# Unified Citizen Identity System Using Blockchain

Sri Sai Abhishake Gopal Dasari

srisaiabhishakegopald@acm.org

**Abstract**

The citizenship identities of a nation's occupants enable the state to identify and authenticate them unquestionably. These documents help individual in recognizing themselves and to profit from the rights and advantages given to them by the legislature or the constitution of the land. There are problems in traditional way of issuance f these identities and many hurdles that impede people from getting their benefits or exercising their rights. These paper-based identities can be forged easily and are hard to authenticate at various civil end points. There are reports of identity thefts. In this paper we are discussing how Blockchain can be employed to overcome these problems and makes these identities confidential, immutable and secured. The blockchain technology can help the governing bodies in maintaining and verifying these identities in a quick manner with less chance for human errors, meaning more reach for government plans and aid

## 1. Introduction

The citizenship records of a country is made out of fundamental distinguishing proof subtleties of its residents that empower the state to distinguish them unmistakably. The identity of the citizens will be established and authorized by identity issuing whose working is controlled by the legislature of the particular nation. This documents or identities help individual in identifying themselves and to avail the rights and benefits provide to them by the government or the constitution of the land. Issuance of multiple identity documents for various purpose makes the data redundant and updating of the data across all the departments takes considerable amount of time and resources. The problem with improper recording of data can lead to case where certain individual has to face severe problems. Even the forging and counterfeiting in the existing identity model of identification has become easy as there is no standard procedure for the verifying the identity. The amount of data one needs to provide to avail a particular service or to exercise a right in form of paper which is hard to verify if it is damaged.

Identity thefts can happen in different forms. Numerous unfortunate victims reported episodes in which their identities and personal information were compromised through various sources.

*Email address:* srisaiabhishakegopald@acm.org (Sri Sai Abhishake Gopal Dasari)

In this paper we would like to discuss the usage of Blockchain technology to address these problems. Blockchain is the collected storage of the data in interconnected blocks which is secured using the cryptographic methods. This technology makes the data immutable, transparent and the data cannot be erased from the storage making it safe and trusted. With Blockchain the individual will have the control over with whom he/she is sharing his/her data other than the authorized government representatives. There can be no intrusion or tampering of the data. This makes inclusion of wrong data impossible. With blockchain and smart contracts in play the system of identity management and verification runs smooth without human intervention and the errors that comes with it.

**Background**

Record keeping of identities of citizens of a nation is done by every country. The citizens avail the benefits and rights by using these government authorized identity documents like AADHAAR[1], PAN[2], Passport, driving license, Voter Id etc. each of these identities are issued and managed by a government body like UIDAI [3], ECI [4]etc. But, these traditional ways of identification have their own shortcomings. They can be damaged, forged, compromised and can be changed in some instances. In some cases, they are not enough to establish a person's identity. Across the world approximately more than one billion people don't have a control over their own identity, making them vulnerable for many hardships, these people cannot exercise their vote, open a bank account or to find a job. These people cannot contribute or be part of the thriving economy. Even the people with proper identity doesn't have a complete ownership of their identity as the third parties that have their data are prone to cyber-attacks and intrusion making them vulnerable as well.

**Blockchain**

Blockchain technology is a recent development of secure method of data storage and record keeping without centralized authority. From a data management standpoint, a blockchain is a distributed database that logs a progressing list of transaction records by consolidating them into an ordered chain of blocks.

A blockchain is formed with the purpose of solving a common business problem that all participants who potentially want to join the network are facing. All stakeholders have their own digital signatures through which identities are established. A transaction is processed and accepted only the after consensus is achieved, and once validated, blocks are sent to all the endpoints or participants, who verify and add it to their chain of blocks, thus forming the blockchain.

Table 1. Types of Blockchain and their properties

Public permissioned blockchain which is a combination of Public and Permissioned type of blockchain, is a suitable type for our use case of citizen identity as it values immutability and efficiency over anonymity because every participant of the blockchain network should be verified by the authorities. To protect the an individual's data from exposure to others we would have access control where only the individual and an authorized governing body have control over their data.

Blockchain can address these problems. Blockchain based unified identity helps in distributing the control over an individual's data between that individual and the government over what, when, where and with whom they are sharing the individual's data. Protecting the privacy and security of the personal details while sharing only the required details with the government agent or organization. The blockchain can resolve the issues like forging, inaccessibility and compromising of identities. Analysis suggest that blockchain technology based identity also help an individual in many ways where he or she can monetize his personal data by opting to share his/her details with a research organizations, private mass surveyors as data is very valuable these days and it is estimated that 60% of global GDP will be digitized within next 2 years with which the value for personal data will increase. Blockchain is the way forward for the government to establish an immutable and to curb red tape.

Table 1:

| Type | Anonym-ity | Transparen-cy | Immutabili-ty | Effi-ciency | Confidential-ity |
|---|---|---|---|---|---|
| Public | Yes | Yes | Yes | No | Low |
| Permis-sioned | No | No | Yes | Yes | Medium |
| Private | No | No | No | Yes | Very high |
| Consortium | No | Partial | Yes | Yes | High |

Immutability mean that the data once written should not be changed. In a blockchain the immutability is achieved by employing hashing, consensus mechanism, and the distributed nature of the blockchain. If an intruder can edit his copy of block it will be evident once the block synchronizes with the entire network making the network robust and tamper-evident.

## 2. System

**Interoperability of Blockchain**

One of the best features of blockchain based identity is its interoperability between other networks. Modern day every private and public organizations are moving towards blockchain by establishing consortiums. The identity network help in sharing and fetching the required details with networks. Like fetching a citizen's medical details from the Hospitals consortium or sharing the PAN details with the bank consortium or to a specific bank making the process fast and secure.

*2.1.*

*2.1.1. Smart Contract*

A smart contract is a digital version of a contract that runs on blockchain end nodes and it runs when the certain conditions are met. The goal of smart contracts is to minimize the human intervention.

The smart contract is built on following steps:

- Understanding the problem of the use case

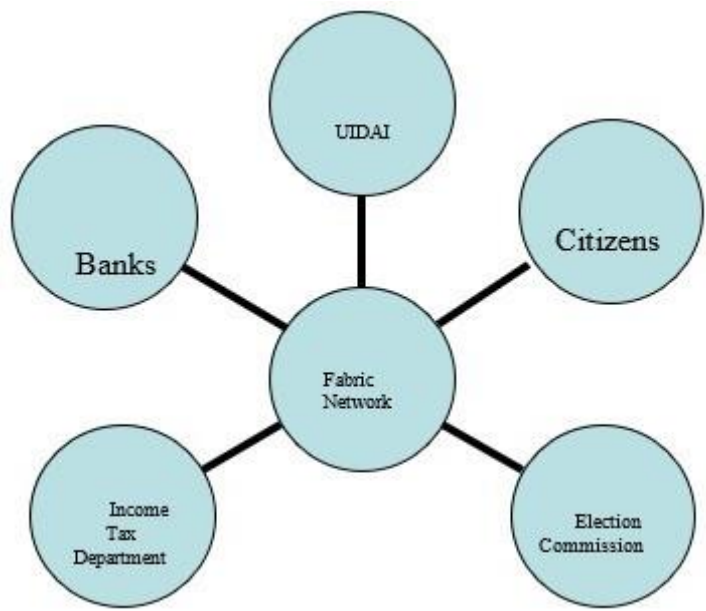

Figure 1: Interoperability of Blockchain making the numerous tasks easy

- Define the properties of network and its stakeholders in smart contract.

- Assessing and adding the functionalities of involved parties of the smart contract.

- Investigative the possible events in contract and conditions that triggered them.

- Evaluation of the smart contact.

As the Blockchain is an append-only decentralized ledger, the smart contracts working through it can just include new transactions on to it. The transactions that happen over the Blockchain can easily verifiable. But then, the most resonating quality of smart contracts is that they have totally minimizes the requirement of trusted third-parties in this framework. As these contracts are self-executing and self-governing in nature and are free from malicious manipulations, they can be utilized to create trustless frameworks which run over an agreement based democratized convention.

**Proposed System**

Here we would like to propose the blockchain based Hyperledger technology as the solution for unified citizen identification, where distributed ledger technology (DLT) will be used. DLT is a system that stores all the information in a decentralized form, making it immutable by any central party (unless authorized). All the data of the network and the changes that have occurred are stored in a ledger, and each participant maintains their own copy of the ledger. A transaction is reflected in the network by consensus, and this makes

*January 17, 2021*

the system secure. Hyperledger Fabric is an open-source framework that can be used to build our use case on permissioned and private blockchains. It is an open-source project based on DLT.

One of the important features of the hyperledger fabric that makes is feasible for our system is its component of Channel.

The key features in Fabric network channel are:
- Channels compartmentalize the data of the network.
- In a Fabric network, the peers can form any number of channels amongst them.
- In a Fabric network, as there can be any number of channels this implies the fact that one peer can be a part of more than just a single channel.

The roles in the proposed system are as follows:

**Client:** It is an application that acts on behalf of any stakeholder in the network. In our use case, authorized personnel from Government body like UIDAI, EIC (Election commission of India) etc., Private organizations like bank or a medical institute and the citizens are the stakeholders, who use the application to access the different functions on the network.

**Peers:** These are the computing resources that are owned by the members of the organizations in the network. Each member can have one or more peers in the network. In the Citizen Identity network, each government bodies, and organizations contribute one or more peer each in the network, and a regulatory body also contributes its peers in the network.

These peers can further have two roles as follows:

**Committer:** It takes the blocks that have been advertised by the ordering service and puts them on its ledger. Every peer that is part of the network will be a committer for a transaction as long as it belongs to the channel where the transaction was initiated.

**Endorser:** The job of the endorser is to take an incoming transaction from the application, simulate the transaction and pass it to the ordering service. A particular peer can be an endorser in the network if it has a chaincode (smart contract) installed on it.

**Ordering service:** An ordering service puts the transactions in the right sequence of their occurrence; converts these transactions into a block and then disseminates this block in the network

The consensus inside Fabric is a three-step process:

Endorsement: Every time a transaction is initiated by any of the peers of the network, that transaction is first endorsed by the peers of the network.

Ordering: Once endorsed, the transaction moves to order. Ordering decides which transactions are put on the block and in what sequence the blocks will be put onto to the ledger of the different peers.

Validation: After the blocks have been ordered in a particular sequence by the ordering service, before being committed on to the ledger by each of the peers, they go through a process of validation.

**The components of our Hyperledger Fabric network are follows:**

Assets: An Asset in Hyperledger Fabric is an important entity that is stored on the ledger. They are stored in the form of key-value pairs. For example, in the Citizen

Identity problem, the Citizen details are the asset in a key-value pair and the representation of the same is as follows:

```
1  NewCitizenObject=
2  {
3  "AADHAR Number" : "12345678911"
4  ``Name'': ``ABC XYZ''
5  ``DOB'': ``DD/MM/YYYY''
6  ``Father Name'': ``XYZ ABC''
7  ``Vote'': ``Not Eligible'',
8  ``PAN'': ``ABC1234P''
9  ``Accounts'':{``SBI00082'':''12345678901''},
10 ``Phone'': 91XXXXXXXX,
11 ``Current State'' : ``Andhra Pradesh'',
12 ``pincode'': 522309,
13 ``Address'': ``DEF street, Door No: 1-1-1, KLM Apartment''
14
15 };
```

Code Snippet 1: Example of a citizen asset.

**Transaction:** A transaction is responsible for the change in the state of the asset. Suppose the person reached an eligible age for voting. This change in a value of asset is referred to as the transaction. The version of the asset "Citizen" will now be updated to version="1" and the vote will be changed from "Not Eligible" to "Eligible".

```
1  {
2  "AADHAR" : "12345678911"
3  ``Name'': ``ABC XYZ''
4  ``DOB'': ``DD/MM/YYYY''
5  ``Father Name'': ``XYZ ABC''
6  ``Vote'': ``Eligible'',
7  ``PAN'': ``ABC1234P''
8  ``Accounts'':{``SBI00082'':''12345678901''},
9  ``Phone'': 91XXXXXXXX,
10 ``Current State'' : ``Andhra Pradesh'',
11 ``pincode'': 522309,
12 ``Address'': ``DEF street, Door No: 1-1-1, KLM Apartment''
13 }
```

Code Snippet 2: Updated citizen asset after the transaction

Ledger: The information about an asset or a transaction of the network is stored in the data structures. The ledger is a data structure which stores all the changes in the assets and transaction. It also stores the current value of the assets as well as the history of the transactions which lead to the current state of the asset.

```
1  Const companyKey = ctx.stub.createCompositeKey
2  ('org.citizen-network.citizennet.citizen', [AADHAR Number])
3  let citizenBuffer =Buffer.from (JSON.stringify
4  (newCitizenObject));
5  await ctx.stub.putState(citizenKey,citizenBuffer);
```
Code Snippet 3: Writing the citizen asset object on to the ledger

World State: The world state is the data structure which stores the current state of the asset. This helps in retrieving the currents state of the citizen data but the Hyperledger provides function to retrieve the historical transactions meaning we can see all the states that an asset has been through.

```
1  async viewaadhar (ctx,AADHAR){
2  const citizenKey=ctx.stub.createCompositeKey ('org.citizen-
       network.citizennet.citizen',[ AADHAR]);
3
4  //getting the current state of citizen
5
6  let citizenBuffer = await
7  ctx.stub.getState(citizenKey).catch(err=>console.log(err)
8  );
9   return JSON.parse(citizenBuffer.toString());
10
11 }
```
Code Snippet 4: Getting the current state of the citizen asset.

```
1  async viewHistory (ctx, AADHAR) { //getting the drug key
2  const citizenKey = ctx.stub.createCompositeKey ('org.citizen-
       network.citizennet.citizen',[ AADHAR]);
3  //getting the history for the key let iterator = await ctx.
       stub.getHistoryForKey(citizenKey); let result = []; let res
        = await iterator.next(); while (!res.done) { if (res.value
       ) {
4  const obj = JSON.parse(res.value.value.toString('utf8'));
       result.push(obj);
5
6   }
7  res = await iterator.next();
8  } await iterator.close(); return result;
9  }
```
Code Snippet 5: Getting the historical states of the citizen asset.

*January 17, 2021*

**Chaincode:** Chaincode is the basic logic for performing and validating the transactions on the fabric network, it enables users to perform transactions on the network's ledger to update and work with assets on it. Chaincode is invoked by triggering transactions that result in a change of the state of the ledger.Chaincode for the citizen network can minimize the human interaction on the all ends.

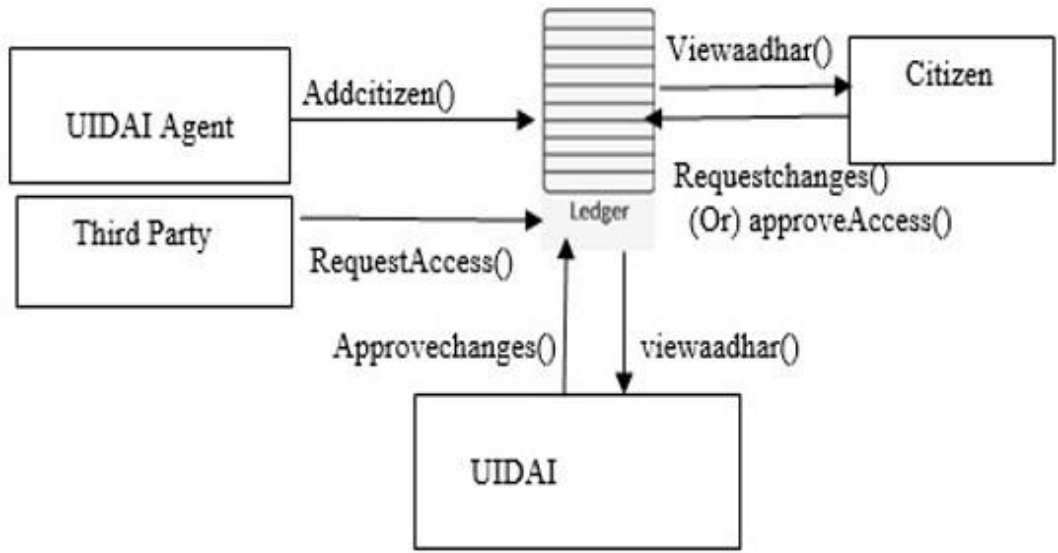

Figure 2: System Architecture

The chaincode is a collection of smart contracts that works to eliminate the errors that will happen due to human interaction. The above picture shows a simple chaincode where different stakeholders interact with the hyperledger using the chaincode functions. Only UIDAI can make changes to the hyperledger where the citizen can request for changes. The third parties can only get access to an individual's aadhar with his/her approval. The chaincode can also help in automating various government or private organizations tasks.

```
1  async updateVoteEligibility(ctx,AADHAR,age) { //fetching the
      citizen details
2  const citizenKey = ctx.stub.createCompositeKey ('org.citizen-
      network.citizennet.citizen',[ AADHAR]);   let citizenBuffer=
       await ctx.stub.getState(citizenKey).catch(err => console.
      log(err)); const citizen= JSON.parse(citizenBuffer.toString
      ());
3  var status = "Eligible";   if(age>=18){ newCitizenObject ={  "
      AADHAR" : "12345678911"   ''Name'': ''ABC XYZ''   ''DOB'': ''
      DD/MM/YYYY''   ''Father Name'': ''XYZ ABC''   ''Vote'':
      status,   ''PAN'': ''ABC1234P''   ''Accounts'':{''SBI00082
```

```
        '':''12345678901''},   ''Phone'': 91XXXXXXXX,    ''Current
        State'' : ''Andhra Pradesh'',   ''pincode'': 522309,   ''
        Address'': ''DEF street, Door No: 1-1-1, KLM Apartment''
        ''VoterID'': xxxxx   }; }
4              //Writing back let citizenBuffer =Buffer.from (
                  JSON.stringify(newCitizenObject));
5   await ctx.stub.putState(citizenKey,citizenBuffer); }
```
Code Snippet 6:

This chaincode once the user (citizen ) raises a request to vote from is client side UI, it compares the present date with the DOB of the person and if the difference is greater than equal to 18 then his voter Id gets generated for the constituency of the pin code of his original address making eligible to cast his vote there.

## 3. Discussion on Scope

The interoperability of the project makes it a unique model which can make tedious amount of work go away and has applications ranging from finance sector to healthcare. Some of those use cases are

- This unified blockchain can help the banks to verify and validate a citizen's financial history with his/her consent to do so. They can get access to their credit score and the chaincode automates the process of loan or credit approval based on the scores and their historical transactions.

- The Model helps in automating a secured and fast way of performing KYC by various private organizations with the citizen's approval and without exchange of a load of documents and paperwork.

- This interoperability nature of this model also helps the in fighting an epidemic or a pandemic to track medical and travel history of citizen by the government authorities and to contain the outbreaks.

- The blockchain can be made to trigger an alert to the authorities when certain behavior is seen in the data of an individual to alert them and track to take action on the person responsible without manually monitoring them all the time

The unified model of the citizens makes the private data secure manageable while making the authorities to have an eye over malicious activities and to trace them quickly

## 4. Conclusion

In this paper we have discussed the foundation of what we believe can be used in various departments to create a quick and transparent system that would benefit the citizens and

*January 17, 2021*

the government. Blockchain has the potential to address the modern day deman for privacy and transparency.

The model we presented our model for citizen Identity model for India using the hyperledger shows how an individual holds control over his data and how it makes their life hassle free and secure.

## 5. List of abbreviations

1. AADHAR - Aadhaar is a verifiable 12-digit identification number issued by UIDAI to the resident of India for free of cost.
2. PAN - Permanent Account Number
3. UIDAI - Unique Identification Authority of India
4. ECI - Election Commission of India